\documentstyle[12pt,{caption}]{article}
\font\tenbf=cmbx10
\font\tenrm=cmr10
\font\tenit=cmti10
\font\elevenbf=cmbx10 scaled\magstep 1
\font\elevenrm=cmr10 scaled\magstep 1
 1

\renewenvironment{thebibliography}[1]
 { \elevenrm
   \begin{list}{\arabic{enumi}.}
    {\usecounter{enumi} \setlength{\parsep}{0pt}
     \setlength{\itemsep}{3pt} \settowidth{\labelwidth}{#1.}
     \sloppy
    }}{\end{list}}
\newcommand {\ignore}[1]{}

\newcommand{\bc}{\begin{center}}
\newcommand{\ec}{\end{center}}

\def\ifmath#1{\relax\ifmmode #1\else $#1$\fi}
\def\3quarter{{\textstyle{3 \over 4}}}

\def\ra{\rightarrow}

\def\lf{\leaders\hbox to 1em{\hss.\hss}\hfill}

\def\21{$SU(2) \ot U(1)$}
\def\ne{\hbox{$\nu_e$ }}
\def\nm{\hbox{$\nu_\mu$ }}
\def\nt{\hbox{$\nu_\tau$ }}
\def\ns{\hbox{$\nu_{sterile}$ }}
\def\nx{\hbox{$\nu_x$ }}
\def\Nt{\hbox{$N_\tau$ }}
\def\ns{\hbox{$\nu_S$ }}

        \def\etc{\hbox{\it etc. }}
\def\eg{\hbox{\it e.g., }}        
\def\etal{\hbox{\it et al., }}
\def\rh{\hbox{right-handed }}

\def\gau{\hbox{gauge }}

\def\sm{\hbox{standard model }}

\def\neu{\hbox{neutrino }}
\def\sa{\hbox{such as }}

\def\neus{\hbox{neutrinos }}

\def\smp{\hbox{standard model. }}

\def\neusp{\hbox{neutrinos. }}

\def\eq#1{{eq. (\ref{#1})}}

\def\fig#1{{Fig. (\ref{#1})}}
\def\VEV#1{\left\langle #1\right\rangle}

\def\lsim{\raise0.3ex\hbox{$\;<$\kern-0.75em\raise-1.1ex\hbox{$\sim\;$}}}
\def\gsim{\raise0.3ex\hbox{$\;>$\kern-0.75em\raise-1.1ex\hbox{$\sim\;$}}}
\def\bel{\begin{letter}}
\def\eel{\end{letter}}
\def\beq{\begin{equation}}
\def\eeq{\end{equation}}
\def\bef{\begin{figure}}
\def\eef{\end{figure}}
\def\bet{\begin{table}}
\def\eet{\end{table}}
\def\bea{\begin{eqnarray}}
\def\ba{\begin{array}}
\def\ea{\end{array}}
\def\bi{\begin{itemize}}
\def\ei{\end{itemize}}
\def\ben{\begin{enumerate}}
\def\een{\end{enumerate}}
\def\ra{\rightarrow}
\def\ot{\otimes}
\def\eea{\end{eqnarray}}
\def\apj#1#2#3{          {\it Astrophys. J. }{\bf #1} (19#2) #3}

\def\ib#1#2#3{           {\it ibid. }{\bf #1} (19#2) #3}

\def\nps#1#2#3{          {\it Nucl. Phys. B (Proc. Suppl.) }
                         {\bf #1} (19#2) #3}
\def\np#1#2#3{           {\it Nucl. Phys. }{\bf #1} (19#2) #3}
\def\pl#1#2#3{           {\it Phys. Lett. }{\bf #1} (19#2) #3}
\def\pr#1#2#3{           {\it Phys. Rev. }{\bf #1} (19#2) #3}
\def\prep#1#2#3{         {\it Phys. Rep. }{\bf #1} (19#2) #3}
\def\prl#1#2#3{          {\it Phys. Rev. Lett. }{\bf #1} (19#2) #3}
\def\pw#1#2#3{          {\it Particle World }{\bf #1} (19#2) #3}

\def\n.c.#1#2#3{         {\it Nuovo Cim. }{\bf #1} (19#2) #3}
\def\r.n.c.#1#2#3{       {\it Riv. del Nuovo Cim. }{\bf #1} (19#2) #3}
\def\sjnp#1#2#3{         {\it Sov. J. Nucl. Phys. }{\bf #1} (19#2) #3}

\def\zetf#1#2#3{         {\it Z. Eksp. Teor. Fiz. }{\bf #1} (19#2) #3}

\def\mpl#1#2#3{          {\it Mod. Phys. Lett. }{\bf #1} (19#2) #3}

\def\ppnp#1#2#3{           {\it Prog. Part. Nucl. Phys. }{\bf #1} (19#2) #3}

\def\tp{these proceedings}
\def\pc{private communication}

\relax
\begin{document}
\begin{center}{{\tenbf
               RECENT PROGRESS IN THE PHYSICS OF NEUTRINO MASS\\}
\vglue 1.0cm
{\tenrm JOS\'E W. F. VALLE}
\footnote{E-mail VALLE at vm.ci.uv.es or 16444::VALLE}\\
\baselineskip=13pt
{\tenit Instituto de F\'{\i}sica Corpuscular - C.S.I.C.\\
Departament de F\'{\i}sica Te\`orica, Universitat de Val\`encia\\}
\baselineskip=12pt
{\tenit 46100 Burjassot, Val\`encia, SPAIN         }\\
\vglue 0.8cm
{\tenrm ABSTRACT}}
\end{center}
\vglue 0.3cm
{\rightskip=3pc
 \leftskip=3pc
 \tenrm\baselineskip=12pt
 \noindent
After briefly reviewing the present limits on \neu masses,
I summarize the cosmological and astrophysical hints from
dark matter, solar and atmospheric neutrino observations that
suggest neutrinos to be massive.
I then focus on other possible manifestations of the \neu
sector at the laboratory, in particular lepton flavour
violating decays \sa
	$\mu \ra e \gamma$,
	$\mu \ra 3 e $,
	$\tau \ra e \pi^0$,
	$\tau \ra e \gamma$,
as well as two-body decays with the emission
of a superweakly interacting spin zero particle,
called majoron, e.g. in $\mu \ra e + J$, $\tau \ra \ell + J$
($\ell = e,\mu$). All of these decays may occur at
levels consistent with present or planned sensitivities,
without violating any experimental data.
In our examples, the underlying physics may be probed also
at LEP, through related processes, e.g. $Z \ra \Nt \nt$ or
$Z \ra \chi \tau$, where \Nt denotes a neutral heavy
lepton, while $\chi$ denotes the lightest chargino.
Another possible, albeit quite indirect, manifestation
of massive \neus is in Higgs physics. As an example
I discuss the possibility of invisibly decaying Higgs,
a quite generic feature of majoron models where the
lepton number is spontaneously violated close to the
electroweak scale. I give the present limits from
LEP1, as well as the LEP2 expectations. I stress
that LEP1 could certainly have missed the Higgs boson
in these models, thus reopening the possibility that
baryogenesis may be just an electroweak phenomenon.
\vglue 0.6cm}

{\elevenbf\noindent 1. Neutrino Masses: Limits and Hints}
\vglue 0.3cm
\hspace{\parindent}

No solid theoretical principle precludes \neus from
having mass. In fact, most attractive extensions of the
\sm require \neus to be massive \cite{fae}. However, theory is
not capable of predicting the scale of \neu masses any
better than it can fix the masses of the other
quarks and charged leptons, say the muon.

There are several limits on \neu masses that follow
from observation. The laboratory bounds may be
summarized as \cite{PDG92}
\beq
\label{1}
m_{\nu_e} 	\lsim 9.3 \: \rm{eV}, \:\:\:\:\:
m_{\nu_\mu}	\lsim 270 \: \rm{keV}, \:\:\:\:\:
m_{\nu_\tau}	\lsim 31  \: \rm{MeV}
\eeq
These limits follow purely from kinematics
and have therefore the great advantage that they
are the most model-independent of the \neu mass
limits. Note that the limit on the \nt mass may
be substantially improved at a tau factory \cite{jj}.
In addition, there are limits
on neutrino masses that follow from the nonobservation of
neutrino oscillations. I address you to ref.
\cite{granadaosc} for a detailed discussion
and compilation. As opposed to the limits in \eq{1}
\neu oscillation limits are correlated ones, involving
\neu mass differences versus mixing. Thus they rely on
the additional assumption, although quite natural in
\gau theories, that massive \neus do mix.

Apart from the above limits, there is an important
one derived from the non-observation of the
${\beta \beta}_{0\nu}$ nuclear decay process i.e.
the process by which nucleus $(A,Z-2)$ decays to
$(A,Z) + 2 \ e^-$.
This lepton number violating process would arise
via \neu exchange, as shown in \fig{betabeta}.
\bef
\vspace{4cm}
\caption{
Neutrino exchange mechanism for ${\beta \beta}_{0\nu}$ decay.}
\label{betabeta}
\eef
Although highly favoured by phase space
over the usual $2\nu$ mode, the neutrinoless
process proceeds only if the virtual neutrino
is a Majorana particle. The decay amplitude is
proportional to
\beq
\VEV{m} = \sum_{\alpha} {K_{e \alpha}}^2 m_{\alpha}
\label{AVERAGE}
\eeq
where $\alpha$ runs over the light neutrinos.
The non-observation of ${\beta \beta}_{0\nu}$
in $^{76} \rm{Ge}$ and other nuclei leads to the limit \cite{Avignone}
\beq
\label{bb}
\VEV{m} \lsim 1 - 2 \ eV
\eeq
depending on nuclear matrix elements \cite{haxton_granada}.
Even better sensitivity is expected from the upcoming
enriched germanium experiments \cite{Avignone}.
Although rather stringent, the limit in \eq{bb}
is rather model-dependent, and does not apply
when total lepton number is an unbroken symmetry,
as is the case for Dirac \neusp Even if all \neus
are Majorana particles, $\VEV{m}$ may differ substantially
from the true neutrino masses $m_\alpha$ relevant for kinematical
studies, since in \eq{AVERAGE} the contributions of
different neutrino types may interfere destructively,
similarly to what happens in the simplest Dirac \neu case,
where the lepton number symmetry enforces that
$\VEV{m}$ automatically vanishes \cite{QDN}.

The ${\beta \beta}_{0\nu}$ decay process may
also be engendered through the exchange of scalar
bosons, as illustrated in \fig{betabeta_scalar}.
\bef
\vspace{5cm}
\caption{Scalar exchange mechanism for ${\beta \beta}_{0\nu}$ decay.}
\label{betabeta_scalar}
\eef
A simple but essentially rigorous proof
\cite{BOX} shows that, in a gauge theory,
whatever the origin of ${\beta \beta}_{0\nu}$
is, it requires \neus to be Majorana
particles, as illustrated in \fig{box}.
\bef
\vspace{4.5cm}
\caption{${\beta \beta}_{0\nu}$ decay and Majorana neutrinos.}
\label{box}
\eef
Indeed, any generic "black box" mechanism inducing
neutrinoless double beta decay can be closed, by W exchange,
so as to produce a diagram generating a nonzero Majorana
neutrino mass, so the relevant neutrino will,
at some level, be a Majorana particle \cite{BOX}.

Gauge theories may lead to new varieties of
neutrinoless double beta decay involving
the $emission$ of light scalars, such as
the majoron \cite{GGN}
\footnote{A related light scalar
boson $\rho$ should also be emitted.}
\beq
(A,Z-2) \rightarrow (A,Z) + 2 \ e^- + J \:.
\eeq
The emission of such weakly interacting light
scalars would only be detected through their
effect on the $\beta$ spectrum.

The simplest model leading to sizeable majoron
emission in $\beta\beta$ decays involving an
isotriplet majoron \cite{GR} leads to a new
invisible decay mode for the neutral \gau
boson with the emission of light scalars,
\beq
Z \ra \rho + J,
\label{RHOJ}
\eeq
now ruled out by LEP measurements of the
invisible Z width \cite{LEP1}.

However it has been recently shown that a sizeable
majoron-neutrino coupling leading to observable
emission rates in neutrinoless double beta decay
can be reconciled with the LEP results in models
where the majoron is an isosinglet and lepton number
is broken at a low scale \cite{ZU}. An alternative
possibility was discussed in \cite{Burgess93}.
This is specially interesting now in view of the
puzzling features recently hinted in some double
beta decay experiments, which might indicate the
presence of very light scalars \cite{klapdor_wein}.

In addition to laboratory limits, there is a cosmological
bound that follows from avoiding the overabundance of
relic neutrinos \cite{KT}
\beq
\sum_i m_{\nu_i} \lsim 50 \: \rm{eV}
\label{rho1}
\eeq
This limit is also model-dependent, as it only holds
if \neus are stable on cosmological time scales.
There are many ways to make neutrinos decay in such
a way as to avoid \eq{rho1} \cite{fae}. The models rely
on the existence of fast \neu decays involving
the emission of a superweakly interacting spin zero
particle, called majoron,
\beq
\nu_\tau \ra \nu_\mu + J
\label{NUJ}
\eeq
The resulting lifetime can be made sufficiently
short so that neutrino mass values as large as
present laboratory limits are fully consistent
with astrophysics and cosmology. Lifetime estimates
in seesaw type majoron models have been discussed
in ref. \cite{V}. Here I borrow the estimate of
ref. \cite{ROMA}.

Curve C in \fig{ntdecay} shows allowed \nt decay
lifetimes as a function of the \nt mass.
Comparing with the \nt decay lifetime needed
in order to efficiently suppress the relic \nt
contribution one sees that the theoretical lifetimes
can be shorter than required. Moreover, since these
decays are $invisible$, they are consistent with all
astrophysical observations.

If, however, the
universe is to have become matter-dominated by a redshift
of 1000 at the latest (so that fluctuations have grown by
the same factor by today), the \nt lifetime has
to be even shorter \cite{ST}, as indicated by the
dashed line in \fig{ntdecay}. Again, lifetimes below
this line are possible. However, this lifetime limit
is less reliable than the one derived from the critical
density, as there is not yet an established theory for the
formation of structure in the universe.

Recently Steigman and
collaborators have argued that many values of the \nt mass
can be excluded by cosmological big-bang nucleosynthesis,
even when it decays \cite{BBNUTAU}. This, however, still
leaves open a wide region of theoretically interesting
\nt lifetime-mass values for which the searches for
the new phenomena suggested here are meaningful.
\bef
\vspace{9.5cm}
\caption{
Estimated \nt lifetime versus observational limits.
The solid line gives the lifetime required in order
to suppress the relic \nt contribution.
The dashed line is suggested from galaxy formation.}
\label{ntdecay}
\eef
As a result, any effort to improve present limits on
\neu masses is definitely worthwhile. These
include experiments searching for distortions
in the energy distribution of the electrons and
muons coming from decays \sa
$\pi, K \ra e \nu$, $\pi, K \ra \mu \nu$, as
well as in nuclear $\beta$ decays \cite{Deutsch}.

In addition to {\sl limits}, observation also
provides us with some positive {\sl hints} for neutrino masses.
These follow from cosmological, astrophysical
and laboratory observations which I now discuss.

Recent observations from large scale data by the
COBE  satellite, when combined with smaller
scale observations (cluster-cluster correlations)
indicate the need for the existence of a hot
{\sl dark matter} component, with $\Omega_{HDM} \sim 0.3$ \cite{cobe}.
For this the most attractive particle candidate is a massive
\neu, \sa a stable \nt of a few eV mass.
This suggests the possibility of having
observable \ne to \nt or \nm to \nt oscillations in the
laboratory. With good luck the next generation
of experiments \sa CHORUS and NOMAD at CERN
and the P803 experiment proposed at Fermilab
will probe this possibility \cite{chorus}.

Second, the {\sl solar \neu data} collected
up to now by the two high-energy
experiments Homestake and Kamiokande,
as well as by the low-energy data on
pp neutrinos from the SAGE and GALLEX
experiments still pose a persisting puzzle
\cite{Davis,granadasol}.
The astrophysical explanation of the
high energy data data would require
not only too large a drop in the temperature
of the solar core, but would also predict wrongly
the relative degree of suppression observed in
Kamiokande and Homestake \cite{Smirnov_wein}.
For a quantitative analysis see ref.
\cite{NEEDNEWPHYSICS}.
The most attractive way to account for the data
is to assume the existence of \neu conversions
involving very small \neu masses $\lsim 10^{-2}$ eV
(MSW mechanism \cite{MSW}). The region of parameters
allowed by present experiments is illustrated
in \fig{msw} \cite{GALLEX} (for recent analyses,
see ref. \cite{MSWPLOT}). Note that the fits
favour the non-adiabatic over the large mixing
solution, due mostly to the larger reduction of
the $^7 $ Be flux found in the small angle region.
\bef
\vspace{9.5cm}
\caption{Region of \neu oscillation parameters
allowed by experiment}
\label{msw}
\eef

Finally, there are hints for \neu masses from studies involving
{\sl atmospheric neutrinos}. Although the predicted
absolute fluxes of \neus produced
by cosmic-ray interactions in the atmosphere
are uncertain at the 20 \% level, their
ratios are expected to be accurate to within
5 \% \cite{atmsasso}. Present observations of
contained events by Kamiokande and IMB, as well
as preliminary Soudan 2 data indicate
the existence a muon deficit \cite{atmsasso,atm}.
Such deficit would suggest the existence
of \neu oscillations of the type
\nm to \nx, where \nx is \ne , \nt or \ns,
a sterile neutrino.
However, no deficit has been established
in the studies of up-going muons performed
by Baksan and IMB \cite{atmsasso}. Recent
analyses of total fluxes at IMB combined
with the studies of the stopping/passing
muon ratio can be used to severely constrain
the oscillation parameters, apparently
excluding oscillations of \nm to \nt
with maximal mixing, as expected in some
theoretical models. Similar analyses have
also been performed for the case of \nm to \ns
as well as \nm to \ne channels, where
matter effects play an important role
\cite{lipari}.

Taken at face value, the above astrophysical and cosmological
observations suggest an interesting theoretical puzzle,
if one insists in trying to account for all three observations
on solar, dark matter and atmospheric \neus within
a consistent theory. Indeed, it is not possible to reconcile
these three observations simultaneously in a world with just
the three known \neus. In order to fit all
the data a fourth \neu species is needed
and, from the LEP data on the invisible Z width,
we know that such fourth \neu must be of the sterile
type \ns.

Two basic schemes have been suggested in
which the \ns either lies at the dark matter scale
(heavy \ns \cite{DARK92}) or, alternatively, at the
MSW scale (light \ns \cite{DARK92B}).
In the first case the atmospheric
\neu puzzle is explained by \nm to \ns oscillations,
while in the second it is explained by \nm to \nt
oscillations. Correspondingly, the deficit of
solar \neus is explained in the first case
by \ne to \nt oscillations, while in the second
it is explained by \ne to \ns oscillations. In
the latter case the limits from big-bang
nucleosynthesis can be used to single out the
nonadiabatic solution uniquely. In both cases
it is possible to fit all observations together.
However, in the heavy \ns case there is a clash
with the bounds from nucleosynthesis while, in the
case of light \ns, since the mixing angle characterizing
the \nm to \nt oscillations is nearly maximal, the
solution is in apparent conflict with the most
recent combined analyses.

In short, \neu masses, besides being suggested
by theory, seem to be required to fit present
astrophysical and cosmological observations.

How could they manifest themselves at the
laboratory?

\vglue 0.6cm
{\elevenbf\noindent 2. Rare Decays: $\mu$, $\tau$ and $Z$}
\vglue 0.3cm
\hspace{\parindent}

Many lepton flavour violating (LFV) decays \sa
$\mu \ra e \gamma$, which are exactly forbidden
in the standard model, can exist if \neus are
massive and, in some cases, even if not!
The possible detection of such decays, like that
of \neu oscillations, would definitely signal new
physics, closely related with the properties
of the neutrinos and the leptonic weak interaction.

Neutrino masses can be induced in many ways
without the need to introduce any new lepton,
such as \rh neutrinos, and no new superheavy
mass scales, much larger than that of the
electroweak theory. In this simplest case
there can be large rates for lepton flavour
violating decays. This is the typical situation in many models
with radiative mass generation \cite{zee,Babu88}.
For example, in the models proposed to
reconcile present hints for \neu masses
there can be large rates for lepton flavour violating
muon decays $\mu \ra e \gamma$ and $\mu \ra 3 e$,
or for the corresponding tau decays.
The first may easily lie within the
present experimental sensitivities, while
the latter may lie at the level $10^{-4}$
to $10^{-6}$ in branching ratio \cite{DARK92,DARK92B}.

Alternatively, one may assume that \neu masses arise
at the tree level, due to the exchange of some
hypothetical neutral heavy leptons (NHLS)
\cite{fae}. In the simplest models of seesaw
type \cite{GRS} the NHLS are superheavy, unless
fine-tunings are allowed \cite{Buch92}. However, in other
variants \cite{SST} this is not the case.
While in the minimal case the expected rate for
LFV processes is expected to be low, due to
limits on \neu masses, in other models
\cite{BER}-\cite{CERN} this suppression need
not be present.
As shown in \fig{mpla}, taken from ref. \cite{3E},
present constraints on weak universality violation
allow for decay branching ratios as large as
${\cal O} (10^{-6})$ for some of these processes.
The solid line corresponds to the attainable branching for the
decay $l \ra l_i l_j^+l_j^-$, the dashed line
corresponds to $\tau \ra \pi^0 l_i$, the dotted line
to $\tau \ra \eta l_i$ and the dash-dotted line to
$\tau \ra l_i \gamma$. Here $l_i$ denotes $e$ or $\mu$
and all possible final-state leptons have been summed
over in each case. As one can see, the most favorable
of all the $\tau$ decay channels are $\tau \ra e \gamma$
and $\tau \ra e \pi^0$, the first being dominant for lower
NHL masses in the 100 GeV to 10 TeV range.
\bef
\vspace{8.2cm}
\caption{Maximum estimated branching ratios for
lepton flavour violating $\tau$ decays consistent
with lepton universality. These processes may
occur even if \neus are strictly massless.}
\label{mpla}
\eef

In addition to these rare decays, there can be
others, more closely related to the masses of the
neutrinos. In any model where \neu masses arise from
the spontaneous violation of an ungauged lepton number
symmetry there is, as a result, a physical Goldstone
boson, called majoron. In order for its existence to be
consistent with the measurements of the invisible $Z$
decay width at LEP, the majoron must be a singlet under
the \21 \gau symmetry. Although the original majoron
proposal was made in the framework of the minimal seesaw
model, and required the introduction of a relatively
high energy scale associated to the mass of the \rh
\neus \cite{CMP}, there is no need for this at all.
There is a vast class of models where the lepton number
is broken close to the weak scale and which can produce
a new class of lepton flavour violating decays.
These include single majoron emission in $\mu$
and $\tau$ decays, which would be "seen" as
bumps in the final lepton energy spectrum,
at half of the parent lepton mass in its rest frame.

I will now briefly mention, as an example of this
situation, the models where supersymmetry is realized
in a \21 context in such a way that R parity is broken
spontaneously close to the weak scale \cite{MASI_pot3}.
This model (called RPSUSY, for short) contains a majoron,
as the spontaneous
violation of R parity requires the spontaneous violation
of the continuous lepton number symmetry.
This majoron is consistent with LEP, as it is mostly
an \21 singlet. Its existence leads to single majoron
emitting $\tau$ decays, whose allowed emission rates
have been determined
by varying the relevant parameters over reasonable
ranges and imposing the observational constraints.
The result is illustrated in \fig{tauej}, taken
from ref. \cite{NPBTAU}
\footnote{The curves are for different values of
$\tan\beta$, a SUSY model parameter expected to lie
between one and the top-bottom quark mass ratio.}.
One sees that these majoron emitting $\tau$ decay modes
may occur at a level $10^{-4}$  in branching ratio,
which lies close to the present limit \cite{SINGLE}.
An important role is played by
constraints related to flavour and/or
total lepton number violating processes \sa
those arising from negative neutrino oscillation
and neutrinoless double $\beta$ decay searches,
as well as from the failure to observe secondary
peaks in weak decays, e.g. $\pi, K \ra \ell \nu$,
with $\ell = e$ or $\mu$.
\bef
\vspace{8cm}
\caption{Branching ratios for lepton flavour
violating $\tau$ decays with majoron emission
consistent with observation in the RPSUSY model.}
\label{tauej}
\eef
Similar results hold for the case of the decay $ \tau \ra \mu + J$,
in which case the attainable branching ratio may even exceed $10^{-3}$.
For muon decays the branching ratios may reach $10^{-6}$ \cite{NPBTAU}.

The sensitivities of the planned tau and B factories
to these decay modes have been recently studied and
are summarized in table 1, from ref. \cite{TTTAU}.
\begin{table}
\begin{center}
\caption{Attainable limits for the branching ratios of different $\tau$
decays in a $\tau$-charm Factory and a B Factory for one year run.}
\begin{displaymath}
\begin{array}{|c|c|c|}
\hline
\mbox{channel} & \mbox{tcF} & \mbox{BF} \\
\hline \hline
\tau\rightarrow e J &
\begin{array} {ccc}   & 10^{-5}  & \mbox{(standard-optics)}\\
                      & 10^{-6}  & \mbox{  (monochromator)}
\end{array}
&  5\times 10^{-3}   \\[0.3cm]
\hline
\tau\rightarrow \mu  J  &
\begin{array} {ccc}   & 10^{-3}  & \mbox{      }\\
                      & 10^{-4}  & \mbox{(RICH)}
\end{array}
&   5\times 10^{-3}  \\[0.3cm]
\hline
\begin{array} {c}
\tau\rightarrow e \gamma \\
\tau\rightarrow \mu \gamma
\end{array}
 &  10^{-7} & 10^{-6}\\[0.3cm]
\hline
\tau\rightarrow \mu \mu\mu     &  &  \\
\tau\rightarrow \mu e e     & 10^{-7}   &10^{-7}  \\
\tau\rightarrow  e  \mu\mu     &     &    \\
\tau\rightarrow  e  e e     &    &    \\[0.3cm]
\hline
\end{array}
\end{displaymath}
\end{center}
\end{table}
One sees that these sensitivities compare well
with the theoretical expectations, summarized
in table 2.
\begin{table}
\begin{center}
\caption{A summary of estimated rare $\tau$ decay
branching ratios consistent with observation.
The processes in the upper box involve majoron emission,
while the ones below are induced by neutral heavy leptons.}.
\begin{displaymath}
\begin{array}{|c|cr|}
\hline
\mbox{channel} & \mbox{strength} & \mbox{} \\
\hline
\tau \ra \mu + J &  \sim 10^{-3} & \\
\tau \ra e + J &  \sim 10^{-4} & \\
\hline
\tau \ra e \gamma ,\mu \gamma &  \sim 10^{-6} & \\
\tau \ra e \pi^0 ,\mu \pi^0 &  \sim 10^{-6} & \\
\tau \ra e \eta^0 ,\mu \eta^0 &  \sim 10^{-6} - 10^{-7} & \\
\tau \ra 3e , 3 \mu , \mu \mu e, \etc &  \sim 10^{-6} - 10^{-7} & \\
\hline
\end{array}
\end{displaymath}
\end{center}
\end{table}
In particular, the planned tau factories
are quite sensitive to the majoron emitting
decays, especially in schemes with monocromator.

The physics of rare $Z$ decays beautifully
complements what can be learned from the
study of rare LFV $\tau$ lepton decays.
For example, the exchange of isosinglet
neutral heavy leptons can induce new $Z$
decays \sa
\footnote{There may also be CP violation in lepton
sector, even when the known \neus are
strictly massless \cite{CP1}. The corresponding
decay asymmetries can be of order unity with respect
to the corresponding LFV decays \cite{CP2}. },
\beq
\begin{array}{lr}
Z \ra e + \tau \: , & Z \ra N_{\tau} + \nu_{\tau}
\end{array}
\eeq
Taking into account the constraints on the parameters
describing the leptonic weak interaction one can
estimate the attainable values for these branching
ratios given in table 3.
\begin{table}
\begin{center}
\caption{Maximum branching ratios for rare $Z$
decays, where \Nt denotes an isosinglet
neutral heavy lepton, while $\chi$ denotes
the lightest chargino (electrically charged
SUSY fermion) and $\chi^0$ is the lightest neutralino.}
\begin{displaymath}
\begin{array}{|c|cr|}
\hline
\mbox{channel} & \mbox{strength} & \mbox{} \\
\hline
Z \ra \Nt \nt &  \sim 10^{-3} & \\
Z \ra e \tau &  \sim 10^{-6} - 10^{-7} & \\
Z \ra \mu \tau &  \sim 10^{-7} & \\
\hline
Z \ra \chi \tau &  \sim 6 \times 10^{-5} & \\
Z \ra \chi^0 \nt &  10^{-4} & \\
\hline
\end{array}
\end{displaymath}
\end{center}
\end{table}
The most copious channel is $Z \ra \Nt \nt$,
possible if the \Nt is lighter than the $Z$
\cite{CERN}. This decay follows from the off-diagonal
neutral currents characteristic of models
with doublet and singlet lepton \cite{2227}.
Subsequent \Nt decays would then
give rise to large missing energy events, called
zen-events. There are now good limits on such decays
from the searches for acoplanar jets and lepton pairs from $Z$
decays at LEP \cite{opal}. The method of experimental
analysis is similar to that used in SUSY zen-event
searches.

If $M_{\Nt}>M_Z$ then \Nt can not be directly produced at
LEP1 but can still mediate rare LFV decays \sa $Z \ra e \tau$
or $Z \ra \mu \tau$ through virtual loops. Under realistic
luminosity and experimental resolution assumptions, it is
unlikely that one will be able to see these processes at LEP
\cite{ETAU}.
In any case, there have been dedicated searches for flavour
violation at the $Z$ peak at LEP, and some limits have already
been obtained \cite{opal}. In contrast, the related low energy processes lie
within the expectations of a $\tau$ or $B$ factory.

Another rare $Z$ decay, characteristic of the
RPSUSY model, is the single production of the lightest
chargino \cite{ROMA,RPCHI},
\beq
Z \ra \chi \tau
\eeq
where the lightest chargino mass is assumed to smaller than
the Z mass. As shown in \fig{5}, the allowed branching ratio
lies close to the present LEP sensitivities. The search for
this decay mode is now underway by the L3 collaboration
\cite{Felcini93}.
\bef
\vspace{8.5cm}
\caption{Branching ratios for single
chargino production in $Z$ decays may exceed
$10^{-5}$ above A and $10^{-6}$ above B.}
\label{5}
\eef
Similarly, the lightest neutralino could also be
singly-produced as $Z \ra \chi^0 \nu_\tau$ \cite{ROMA,RPCHI}.
Being unstable due to R parity violation, $\chi^0$ is
not necessarily an origin of events with missing energy,
since some of its decays are into charged particles.
Thus the decay $Z \ra \chi^0 \nu_\tau$ would give rise to
zen events, similar to those of the minimal supersymmetric
standard model (MSSM), but where the missing energy is
carried by the \nt. Another possibility for zen events
in RPSUSY is the usual pair neutralino production process,
where one $\chi^0$ decays visibly and the other invisibly.
The corresponding zen-event rates can be larger than
in the MSSM.
Their origin is also quite different, being tied with
the nonzero value of the \nt mass. Moreover,
although the \nt can be quite massive, the \ne and \nm
have a tiny mass difference, that can be chosen to lie
in the range where resonant \ne to \nm conversions provides
an explanation of solar \neu data. Due to this peculiar
hierarchical pattern, one can go even further, and regard the
rare R parity violating processes as a tool to
probe the physics underlying the solar \neu deficit
in this model \cite{RPMSW}. Indeed, the rates for
such rare decays can be used in order to
discriminate between large and small mixing
angle MSW solutions to the solar \neu problem.
Typically, in the nonadiabatic region of
small mixing one can have larger rare decay branching
ratios, as seen in Fig. 5 of ref. \cite{RPMSW}.

Finally, it is possible to find manifestations of massive
\neus even at the superhigh energies available at
hadron supercolliders LHC/SSC. An example has
been discussed in the RPSUSY model, i.e. the
possible single production of supersymmetric
fermions \cite{RPLHC}.

The above examples illustrate how the search for
rare decays can be a more sensitive probe of \neu
properties than the more direct searches for \neu
masses, and therefore complementary.

\vglue 0.6cm
{\elevenbf\noindent 3. Invisible Higgs Decays }
\vglue 0.3cm
\hspace{\parindent}

Another possible manifestation, admittedly quite indirect,
of models with massive \neus is in the Higgs sector.
There is, in fact, a very wide class of such models
\cite{JoshipuraValle92} where \neu masses are induced
from the spontaneous violation of a global $U(1)$
lepton number symmetry by an \21 singlet vacuum
expectation value $\VEV{\sigma}$. These extensions
of the minimal standard model may naturally explain
the neutrino masses required by present astrophysical
and cosmological observations.

In all of these models $\VEV{\sigma}$ lies close to
the electroweak scale. Unlike the seesaw majoron case
\cite{CMP}, such a low scale for the lepton number
violation is {\sl preferred} since, in these models,
$m_\nu \to 0$ as $\VEV{\sigma} \ra 0$. As a result,
a relatively low value of $\VEV{\sigma}$ is
required in order to obtain naturally small neutrino
masses, either at the tree level or radiatively
\cite{JoshipuraValle92}
\footnote{Another example is provided by the
RPSUSY models \cite{HJJ}.}.

Another cosmological motivation for low-scale
majoron models has been given in ref. \cite{Goran92}.

In such low-scale models the majoron has
significant couplings to the Higgs bosons,
even though its other couplings are suppressed.
This implies that the Higgs boson will decay
with a substantial branching ratio into the
invisible mode \cite{JoshipuraValle92,Joshi92}
\begin{equation}
h \rightarrow J\;+\;J
\label{JJ}
\end{equation}
where $J$ denotes the majoron. The presence of
this invisible Higgs decay channel can affect
the corresponding Higgs mass bounds in an
important way, as well as lead to novel
search strategies at higher energies,
\sa at hadron supercolliders LHC/SSC
\cite{bj,granada,kane}.

The production and subsequent decay of any Higgs boson
which may decay visibly or invisibly involves three independent
parameters: the Higgs boson mass $M_H$, its coupling
strength to the Z, normalized by that of the \sm, call
this factor $\epsilon^2$, and the invisible Higgs boson
decay branching ratio.

The results published by the LEP experiments on the
searches for various exotic channels can be used
in order to determine the regions in parameter space
that are ruled out already. The procedure was described
in \cite{alfonso}.
For each value of the Higgs mass, the lower bound on
$\epsilon^2$, as a function of the branching ratio
$BR(H \rightarrow $ visible) can be calculated. The highest
such bound for $BR(H \rightarrow $ visible) in the range
between 0 and 1, provides the absolute bound
on $\epsilon^2$ as a function of $M_H$.

For a Higgs of low mass (below 30 GeV) decaying to invisible
particles one considers the process $Z \rightarrow H Z^*$,
with $Z^* \rightarrow e^+e^-$   or $Z^* \rightarrow \mu^+\mu^-$
and the combined the results of the LEP experiments on
the search for acoplanar lepton pairs \cite{OPAL91,ALEPH91,L3_91}
which found no candidates in a total sample corresponding to
780.000 hadronic Z decays. The efficiencies for the detection
of the signal range from 20\% at very low Higgs masses to
almost 50\% for $M_H = 25$ GeV.

For higher Higgs masses the rate of the process used above is too
small, and one can use instead the channel $Z  \rightarrow HZ^*$,
$Z^*  \rightarrow q\bar{q}$. Here the results of the searches for
the Standard Model Higgs in the channel $Z  \rightarrow Z^* H_{SM}$,
with $H_{SM} \rightarrow q\bar{q}$ and $Z \rightarrow  \nu \bar{\nu}$,
can be adapted from ref. \cite{Felcini92}. The efficiency of these
searches for an invisible Higgs increases from 25\% at
$M_H = 30$ GeV to about 50\% at $M_H = 50$ GeV.

For visible decays of the Higgs boson its signature is the same
as that of the Standard Model one, and the searches for this particle
can be applied directly. For masses below 12 GeV the results
of a model independent analysis made by the L3 collaboration
can be taken over (ref. \cite{L3_92}).
For masses between 12 and 35 GeV one combines the results from
references  \cite{OPAL91,Felcini92,L3_92}; finally for
masses up to 60 GeV one uses the combined result of all the
four LEP experiments given in reference \cite{Felcini92}.
In all cases the bound on the ratio
$BR(Z \rightarrow ZH)/BR(Z \rightarrow ZH_{SM})$
can be calculated from the quoted sensitivity,
taking into account the background events where they existed.

Proceeding this way the exclusion contours in the plane
$\epsilon^2$ vs. $BR(H \rightarrow $ visible) can be
determined for each particular choice for the Higgs
mass. The two curves corresponding to the searches
for visible and invisible decays can be combined
to give the final bound on $\epsilon^2$,
independently of the value of
$BR(H \rightarrow $ visible) (see Fig. 1
of ref. \cite{alfonso}.)
Repeating this operation for different Higgs boson masses
one can make an exclusion plot, shown in \fig{alfonso2},
of the region in $\epsilon^2$ vs. $M_H$ that is
already excluded by the present LEP analyses,
{\sl irrespective of the mode of Higgs decay},
visible or invisible \cite{alfonso}.
\bef
\vspace{9.5cm}
\caption{Region in the $\epsilon^2$ vs. $m_H$ that can be
excluded by the present LEP1 analyses, independent of the
mode of Higgs decay, visible or invisible (solid curve).
Also shown are the LEP2 extrapolations (dashed).}
\label{alfonso2}
\eef
Finally, one can also determine the additional
range of parameters that can be covered by LEP2
for a total integrated luminosity of 500 pb$^{-1}$
and centre-of-mass energies of 175 GeV and 190 GeV.
This is shown as the dashed and dotted curves in
\fig{alfonso2}. The results on the visible decays
of the Higgs are based on the study of efficiencies
and backgrounds in the search for the Standard
Model Higgs described in reference \cite{Janot92}.
For the invisible decays of the Higgs the only channel
considered was HZ with $Z \rightarrow e^+e^-$   or
$Z \rightarrow \mu^+\mu^-$, giving a signature of two
leptons plus missing transverse momentum. The requirement
that the invariant mass of the two leptons must be close
to the Z mass can kill most of the background from WW and
$\gamma\gamma$ events; the background from
ZZ events with one of the Z decaying to neutrinos is small and the
measurement of the mass recoiling against the two leptons allows
to further reduce it, at least for $M_H$ not too close to $M_Z$.
Hadronic decays of the Z were not considered, since the background
from WW and $We\nu$ events is very large, and b-tagging is much less
useful than in the search for $Z H_{SM}$ with
$Z \rightarrow  \nu \bar{\nu}$, since the $Zb\bar{b}$ branching ratio
is much smaller than $Hb\bar{b}$ in the \smp

The possibility of invisible Higgs decay
is also very interesting from the point of
view of a linear $e^+ e^-$ collider at higher
energy \cite{EE500}. Heavier, intermediate-mass,
Higgs bosons can also be searched at high energy
hadron supercolliders such as LHC/SSC \cite{granada,bj,kane}.
The limits from LEP discussed above should
serve as useful guidance for such future searches.

\vglue 0.6cm
{\elevenbf\noindent 4. Electroweak Baryogenesis?}
\vglue 0.3cm
\hspace{\parindent}

The requirements necessary in order to generate
the cosmological baryon asymmetry were put
forward in a pioneer paper by Sakharov in 1967.
The conditions on the processes responsible for
creating this asymmetry are :
(i) CP-violation, (ii) B violation and
(ii) out of equilibrium conditions \cite{Sakharov67}.
The observation that nonperturbative B violating
electroweak effects \cite{tHooft76} can be substantial
at high temperatures, as expected to hold in the early
universe  opens the possibility of baryogenesis
at the weak scale \cite{Kuz}.
In fact, even if the baryon asymmetry is due to
physics at superhigh scales, it may be washed out
during the electroweak phase transition, as stressed
by Shaposhnikov and others \cite{Kuz,Shaposhnikov88,Linde,Dineetal92a}.
Although the exact details are still the subject of
controversy, there is now a consensus that, for
reasonable choices of the Higgs boson mass, the
transition can be first order, as required.
However, in order to prevent that this baryon
asymmetry be washed out during the phase
transition one requires \cite{Dineetal92a}
\beq
\label{40a}
m_H \lsim 40 \rm{GeV}
\eeq
Unfortunately this upper bound on the Higgs
boson mass required for successful baryogenesis
is in conflict with the recent combined lower
limit obtained from LEP \cite{LEP1}, i.e.
\beq
\label{60}
m_H \geq 60 \rm{GeV}
\eeq

The possibility of completely evading the experimental
Higgs boson mass limit in low-scale majoron models is
vividly illustrated by \fig{alfonso2}. One sees that,
as long as $\epsilon^2$ is sufficiently low (which is
theoretically natural) the Higgs boson could still have
a very low mass. Another drastic example of this same
phenomenon was found "empirically" in the RPSUSY
model of ref. \cite{HJJ}. This observation reopens
the issue of whether baryogenesis can be just
another phenomenon within the electroweak sector
of the fundamental theory \cite{BGLAST}.

Other possible ways whereby models of \neu mass may
avoid the conflict between \eq{40a} and \eq{60} were
also suggested in ref. \cite{BGLAST}. They share in
common the idea that perhaps \neu physics may underlie
the mechanism responsible for generating the baryon
excess. One question which still remains open,
however, is whether these low-scale majoron
models can provide, on their own, a viable
baryogenesis model. For completeness I mention
that there have been other attempts to resolve
the clash between \eq{40a} and \eq{60} by
extending the Higgs sector in ways unrelated to
\neu physics \cite{AndersonHall92_2Higgs}.

\vglue 0.6cm
{\elevenbf\noindent 5. Conclusion}
\vglue 0.3cm
\hspace{\parindent}

Present cosmological and astrophysical observations,
as well as theory, suggest that \neus may be massive.
Existing data do not preclude \neus from being responsible
for a wide variety of measurable implications at the laboratory.
These new phenomena would cover an impressive region of energy,
from weak decays, \sa $\beta$ and double $\beta$ decays,
to \neu oscillations, to rare processes with lepton flavour
violation, from nuclear $\mu e$ conversions and rare muon
decays, to tau decays, to Z decays.
Indeed, it is rather amazing that \neu masses might
affect even the electroweak symmetry breaking sector
in an important way.

The next generation of experiments \sa those
looking for spectral distortions in weak decays,
experiments with enriched germanium looking for
signs of neutrinoless double $\beta$ decays,
\neu oscillation searches sensitive to \nt
as dark matter (CHORUS/NOMAD/P803),
$e^+ e^-$ collisions from ARGUS/CLEO, to
TCF/BMF (tau-charm and B factories), to
LEP; and finally even the next decade hadron
supercolliders LHC/SSC could all be sensitive
to \neu properties!

It is therefore quite worthwhile to keep
pushing on the cosmological and astrophysical
front with solar \neu experiments \sa the gallium-germanium
SAGE and GALLEX experiments, as well as Superkamiokande,
Borexino, Sudbury and the proposed HELLAZ solar \neu project.
The same can be said of the ongoing
studies with atmospheric \neusp

Similarly, a new generation of experiments
capable of more accurately measuring the cosmological
temperature anisotropies at smaller angular scales than
COBE, would be good probes of different models of
structure formation, and presumably shed further
light on the need for hot \neu dark matter.
All such endeavours should be gratifying.
After all, \neus might even be related
to the observed baryon excess!

\vglue 0.5cm
{\elevenbf\noindent 6. Acknowledgements}
\vglue 0.2cm
\hspace{\parindent}
I thank the organizers for the kind invitation
and friendly organization. This work was supported
by DGICYT, under grant number PB92-0084.

\vglue 0.6cm
{\elevenbf\noindent 7. References}
\hspace{\parindent}

\bibliographystyle{ansrt}

\begin{thebibliography}{99}
\bibitem{Burgess93}
C P Burgess and J M Cline, \pl{B298}{93}{141}. This paper
makes wrong criticisms of the preceeding paper, claiming
that the assumed majoron coupling is in conflict with the
\neu oscillation limits. This is clearly incorrect, as
${\beta \beta}_{0\nu}$ may proceed in a one generation model.
\end{thebibliography}

\end{document}